\documentclass{article}
\usepackage{mrs2005,epsfig}
\begin{document} 
\title{GALEX DETECTION OF LYMAN BREAK GALAXIES AT $z \sim 1$ IN THE CDFS}

\author{D. Burgarella$^1$, P. Perez-Gonzalez$^2$, V. Buat$^1$, T.T. Takeuchi$^1$, S. Lauger$^1$, G. Rieke$^2$, O. Ilbert$^3$}
\affil{$^1$ Observatoire Astronomique Marseille Provence / LAM, France, denis.burgarella@oamp.fr; veronique.buat@oamp.fr; tsutomu.takeuchi@oamp.fr; sebastien.lauger@oamp.fr}
\affil{$^2$ Steward Observatory, University of Arizona, USA; pgperez@as.arizona.edu; grieke@as.arizona.edu}
\affil{$^3$ Osservatorio Astronomico di Bologna, Italy; olivier.ilbert1@bo.astro.it}
 
\begin{abstract} 
The first large sample of Lyman Break Galaxies (LBGs) at $z \sim 1$ is selected from $GALEX$ Far-UV and Near-UV images of the Chandra Deep Field South. This multi-$\lambda$ study is based on the wealth of data available in this field : $SPITZER-MIPS$ to estimate total IR luminosities and dust attenuations, redshifts from COMBO-17, EIS and GOODS to check the validity of dropout identifications and GOODS to derive the morphology of about 40 of our LBGs. The advantages of this "nearby" LBG sample are that images reach fainter fluxes and surface brightnesses, allowing a deeper detection at all wavelengths and a better morphological analysis. Moreover, the 24 $\mu m$ fluxes (i.e. 12$ \mu m$ rest-frame) permit a good estimation of dust attenuations and total IR luminosities. The main results are that the vast majority of our LBGs are also Luminous Infrared Galaxies (LIRGs). The morphology of 75 \% of our LBGs is consistent with a disk. Previous estimates of dust attenuations based on the ultraviolet slope are likely to be too large by a factor of $\sim$ 2, which implies that estimated star formation rates were too large by the same amount. This sample in the $z \sim 1$ universe provides us with a high quality reference sample of LBGs.
\end{abstract} 
 
\section{Introduction} 

Lyman Break Galaxies (LBGs) form the main galaxy population observed in the rest-frame ultraviolet (UV) at high redshift. LBGs are similar to local starbursts and it was suggested that these objects are forming stars at a high rate (Pettini et al. 2000; Teplitz et al. 2000). 
Observed colors of LBGs are redder than dust-free star-forming objects. However, the actual amount 
of dust in LBGs and the total star formation rate (SFR) is still insecure because high redshift LBGs are mainly undetected in the FIR / sub-mm range and only small LBG 
samples are available (Chapman et al. 2000; Huang et al. 2005).
The morphology of LBGs is also a matter of debate (see Giavalisco 2002). Early works suggested that LBGs could be ellipsoidal and perhaps the progenitor of ellipticals or
bulges of today galaxies. The problem is that we can hardly detect low surface brightness areas above $z \sim 1$ (Burgarella et al. 2001). However, Papovich et al. (2005) found that the morphology of a NIR-selected sample of galaxies at $z = 1$ have regular and symmetric morphologies while a $z \sim 2.3$ sample present irregular and/or compact morphologies.

Observations in the sub-mm range introduced FIR-bright galaxies which might be similar to (Ultra) Luminous IR galaxies ((U)LIRGs) (Blain et al. 1999) with 
$10^{11} < L_{IR}/L\sun = L(8-1000\mu m) < 10^{12}$ for LIRGs and $10^{12} < L_{IR}/L\sun < 10^{13}$ for ULIRGs that are likely to dominate the Cosmic Infrared Background (Elbaz \& Cesarsky 2003) at high redshift.

The link between LBGs and LIRGs is still an open question. Are they two different classes of 
objects ? Are they two facets of the same population ? Could
we correct UV fluxes for the dust attenuation to recover the full Star Formation Density ?

In this paper, we combine the detection of true (i.e. with a detected break) LBGs at
$z \sim 1$ thanks to $GALEX$ (e.g. Morrissey et al. 2005) with 24 $\mu m$ $SPITZER/MIPS$ (Rieke et al. 2004) data to estimate the total dust emission
and the dust-to-UV flux ratio. We also use GOODS images to analyse the morphology.
A cosmology with $H_0 = 70 km.s^{-1}.Mpc^{-1}$, $\Omega_M = 0.3$ and
$\Omega_{\Lambda}=0.7$ is assumed.

\section{The definition of the LBG sample}

To build the sample, we use $GALEX$ data that are cross-correlated with COMBO-17 to get redshifts and $SPITZER/MIPS$ to get $24-\mu m$ fluxes. We down-select objects that COMBO-17 (Wolf et al. 2004) clearly puts in the "GALAXY" class. We extract sources with redshifts $z \geq 0.9$. We
keep only objects down to GALEX NUV = 24.5 corresponding to GALEX 80 \%
completeness level (456 objects). Then, we select 344 FUV-dropout objects with $FUV-NUV > 2$. The observed FUV and NUV filters are in the same rest-frame wavelength ranges than $G$ and $R$ and this selection
is consistent with high redshift works (e.g. Giavalisco et al. 2002 for a review). Of this list, 90 objects (23.3 \%) have a measured flux above the 80 \% completeness limit of SPITZER 24 $\mu m$ data
(e.g. P\'erez-Gonz\'alez et al. 2005) and 264 (76.7 \%) only have upper limits. This sample
of 344 LBGs in the redshift range $z \geq 0.9$ constitutes the database that we will study
in this work.

%
%
\begin{figure}  
\vspace*{1.25cm}  
\begin{center}
\epsfig{figure=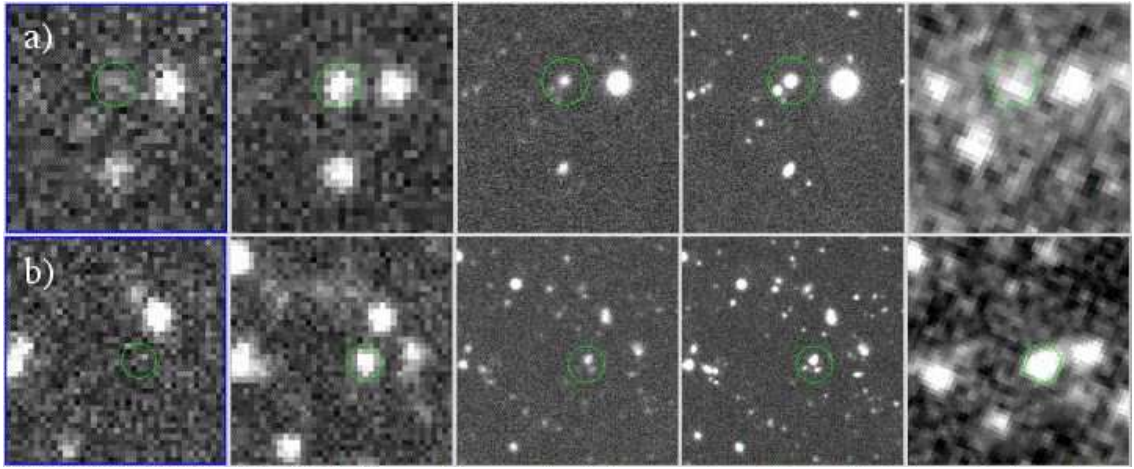,width=15cm}  
\end{center}
\vspace*{0.25cm}  
\caption{Two galaxies from our LBG sample are shown here, from left to right in $GALEX$ FUV and NUV, then $EIS$ B and R and finally $SPITZER-MIPS$ $24 \mu m$ band. For the two galaxies, the leftmost image is below the Lyman break at $z \sim 1$ and the galaxy is not visible. a) a LBG classified as a disk-dominated galaxy (the smaller companion to the left of the LBG is a galaxy at $z \sim 0.3$) and b) as a merger / interacting galaxy.} 
\end{figure} 

This is the first LBG sample in this redshift range that can be used as a basis to
which higher redshift LBGs can be compared. Fig.~1 shows two LBGs at several wavelengths including the two $GALEX$ bands, the B and R EIS bands and the $24 \mu m$ $SPITZER/MIPS$ band.

\section{Main characteristics of Lyman Break Galaxies at $z \sim 1$}

\subsection{Ultraviolet and Infrared Luminosities}

We find a large range of observed (i.e. uncorrected for dust attenuations) FUV luminosities with $9.3 \leq Log (L_{UV}/L\sun) \leq 11.0$ corresponding to $-22.5 \leq M_{1800} \leq -18.0$. The luminosity range (assuming $H_0 = 70$) is wider than either the $z \sim 1$ or the $z \sim 3$ one in Adelberger \& Steidel (2000; A\&S00). Our LBGs reach luminosities as faint as A\&S00 Balmer break sample at $z \sim 1$ but as bright as their LBG sample at $z \sim 3$. Total IR luminosities $L_{IR}$ are estimated from $\lambda f_{24 \mu m}$ (i.e. $12 \mu m$ a t $z = 1$ from Takeuchi et al. 2005). 
In our LBG sample, 20 - 30 \% are UV Luminous Galaxies (UVLGs; Heckman et al. 2005). The IR luminosity range of the sample is about $10.9 < Log (L_{IR})/L\sun < 12.0$,
which means that essentially all LBGs are compatible with a classification into LIRGs. 
An association between UVLGs and LIRGs was already proposed by Burgarella et al. (2005).
We confirm here that UV-luminous galaxies are also IR-luminous objects and extend it to LBGs.

%
%
\begin{figure}  
\vspace*{1.25cm}  
\begin{center}
\epsfig{figure=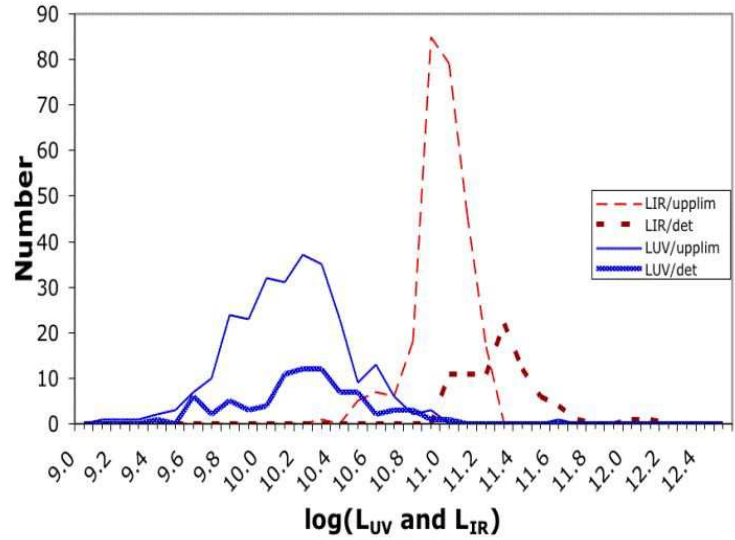,width=10cm}  
\end{center}
\vspace*{0.25cm}  
\caption{The distribution in observed (i.e. non corrected for dust attenuation) luminosity presented here corresponds to UV  luminosities in blue and IR luminosities in red (dashed). Heavy lines are drawn from the population of LBGs with detected counterparts in the $24 \mu m$ MIPS image and thin lines to upper limits only. Note that the cut at low $L_{IR}$ is not sharp because the $83 \mu Jy$ limit used here is not the detection limit by the 80 \% completeness limit. UV luminosities cover the same range for the two samples and reach uncorrected UV luminosities $Log (L_{UV}/L\sun) = 11$. This upper limit is consistent with Adelberger \& Steidel's (2000) range for LBG at $z \sim 3$ but seems inconsistent with the Balmer-break sample at $z \sim 1$.} 
\end{figure} 

\subsection{Morphologies}

The CDFS was observed by GOODS and we can deduce a morphological information in the rest-frame B band. The advantage of observing LBGs at low redshift is obviously to have a better indication
of their true morphology and not only the high surface brightness regions. We used asymmetries and concentrations from Lauger et al. (2005) to build Fig.~2. From $HST$ observations, it was found that LBGs at $z > 2$ can have a wide dispersion of morphology but they generally appear to be smaller, 
irregular-like and compact (Papovich et al. 2005). However, we must be careful in the interpretation 
because of the cosmological dimming (e.g. Burgarella
et al. 2001 but see Papovich et al. 2005). Note, however, that in some rare deep spectroscopic observations (Mooorwood et al. 2000; Pettini et al. 2001), the profile of optical nebular lines suggests the presence of disks in some high-z LBGs.

Fig.~3 leads us
to the three following conclusions: i) all but one LBGs are disk-dominated galaxies, 
ii) one of them might be associated to 
bulge-dominated galaxies within the uncertainties and iii) part of them (21 \%)
are in the top part of the diagram, i.e. with an asymmetry larger than 0.25 and could be
interpreted as mergers.

%
%
\begin{figure}  
\vspace*{1.25cm}  
\begin{center}
\epsfig{figure=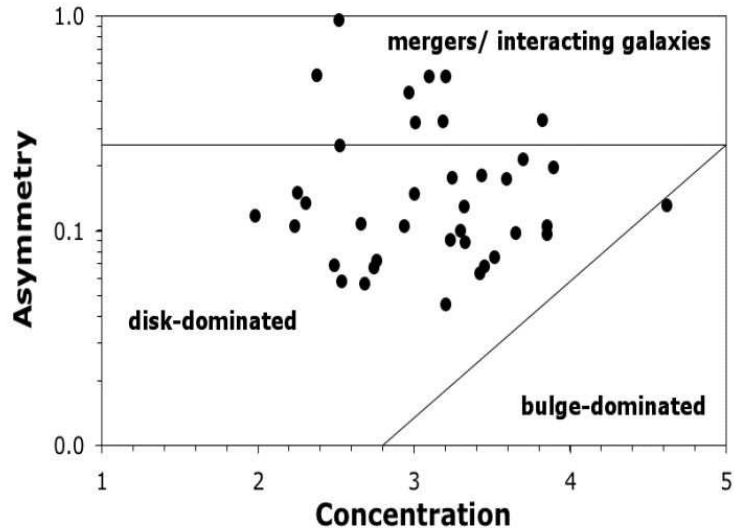,width=10cm}  
\end{center}
\vspace*{0.25cm}  
\caption{The morphology of the LBG sample is estimated from the asymmetry and the concentration (sub-sample drawn from a larger CDFS analysis by Lauger et al. 2005). The line corresponding to the limit between disk-dominated and bulge-dominated galaxies is defined from Lauger et al. (2005) catalogue at z=0. The location in the diagram reflects the morphological type of the galaxies: more asymmetrical LBGs (e.g. mergers) are in the top part of the diagram ($A > 0.25$) while early-type spirals would have $A < 0.1$. The LBG sample is mainly dominated (75 \%) by disk-dominated galaxies and the contribution from mergers amounts to $\sim 21$ \%.} 
\end{figure} 

A number of studies have been devoted to the morphology of galaxy in the redshift range $0.6 \leq z \leq 1.2$ (e.g. Wolf et al. 2005; Lauger et al. 2005; Zheng et al. 2004; Papovich et al. 2005). The figure emerging from these papers is that most of the star formation at $0.6 < z < 1$ resides in disks and about half in spirals. We find that $\sim 21$ \% of the LBGs are likely mergers (e.g. Fig.~1b), $\sim 75$ \% are disks (e.g. Fig.~1a) and only $\sim 3$ \% (i.e. 1 galaxy) is a spheroid. Given that almost all our LBGs can also be classified as LIRGS from their IR luminosity, our classification is also consistent with the LIRG morphology carried out by Zheng et al. (2004). 

\subsection {Ultraviolet Dust Attenuations and Implications on the Cosmic Star Formation Density}

So far, it was difficult to estimate the validity of dust attenuations of LBGs, because we had no clear idea of $L_{IR}$. A\&S00 tried to estimate the $800 \mu m$ fluxes of their LBG sample from the $\beta$ method (e.g. Calzetti et al. 1994) and compared them to observations. But, only the most extreme ones can be detected and used to safely estimate a calibration. In this paper, we use total IR luminosities $L_{IR}$, and $L_{UV}$ to compute the FIR/UV ratio which is calibrated into FUV dust attenuation $A_{FUV}$ (e.g. Burgarella et al. 2005). This method has been shown to provide better dust attenuations than the UV slope $\beta$.

%
%
\begin{figure}  
\vspace*{1.25cm}  
\begin{center}
\epsfig{figure=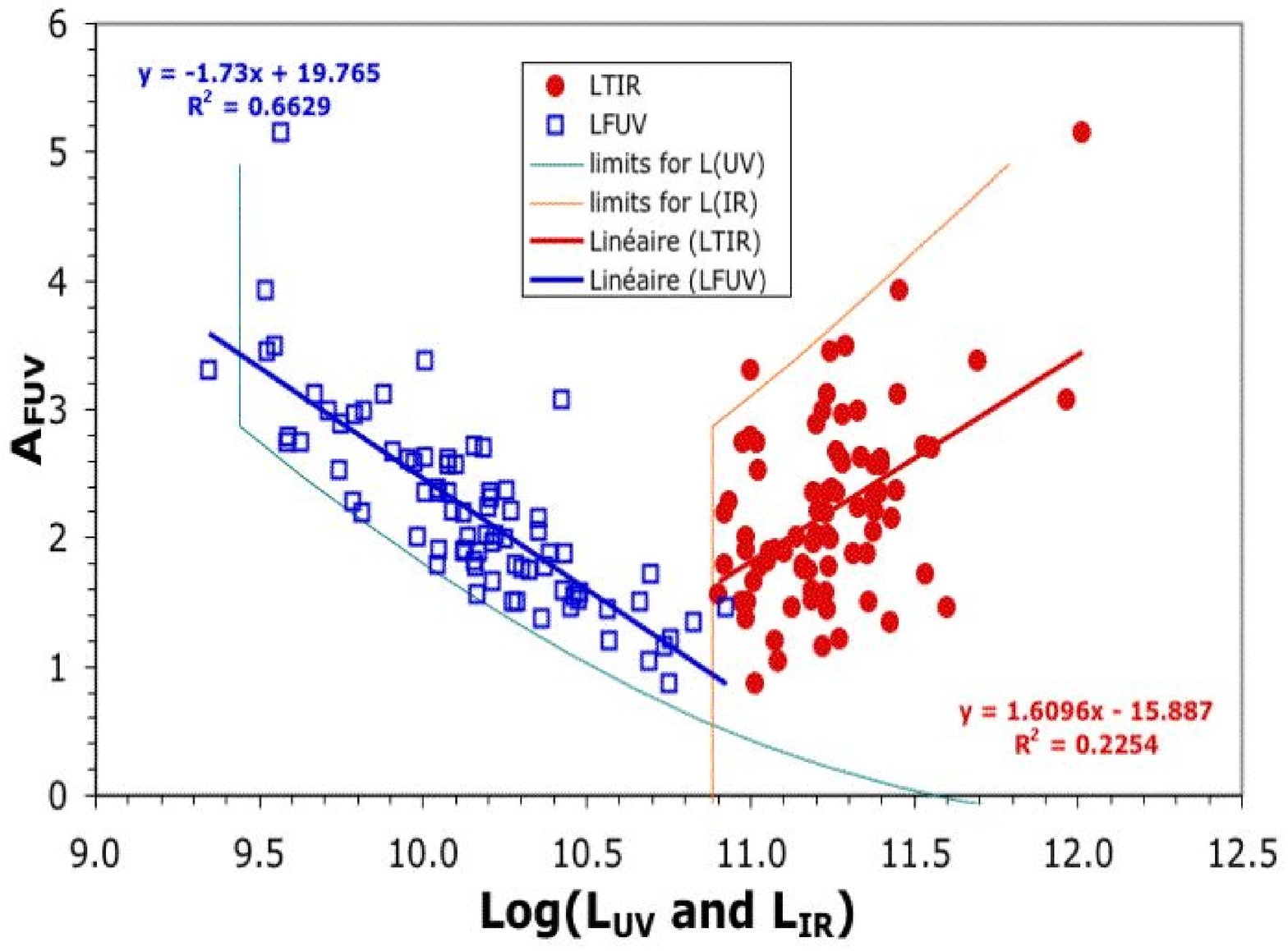,width=8cm}  
\epsfig{figure=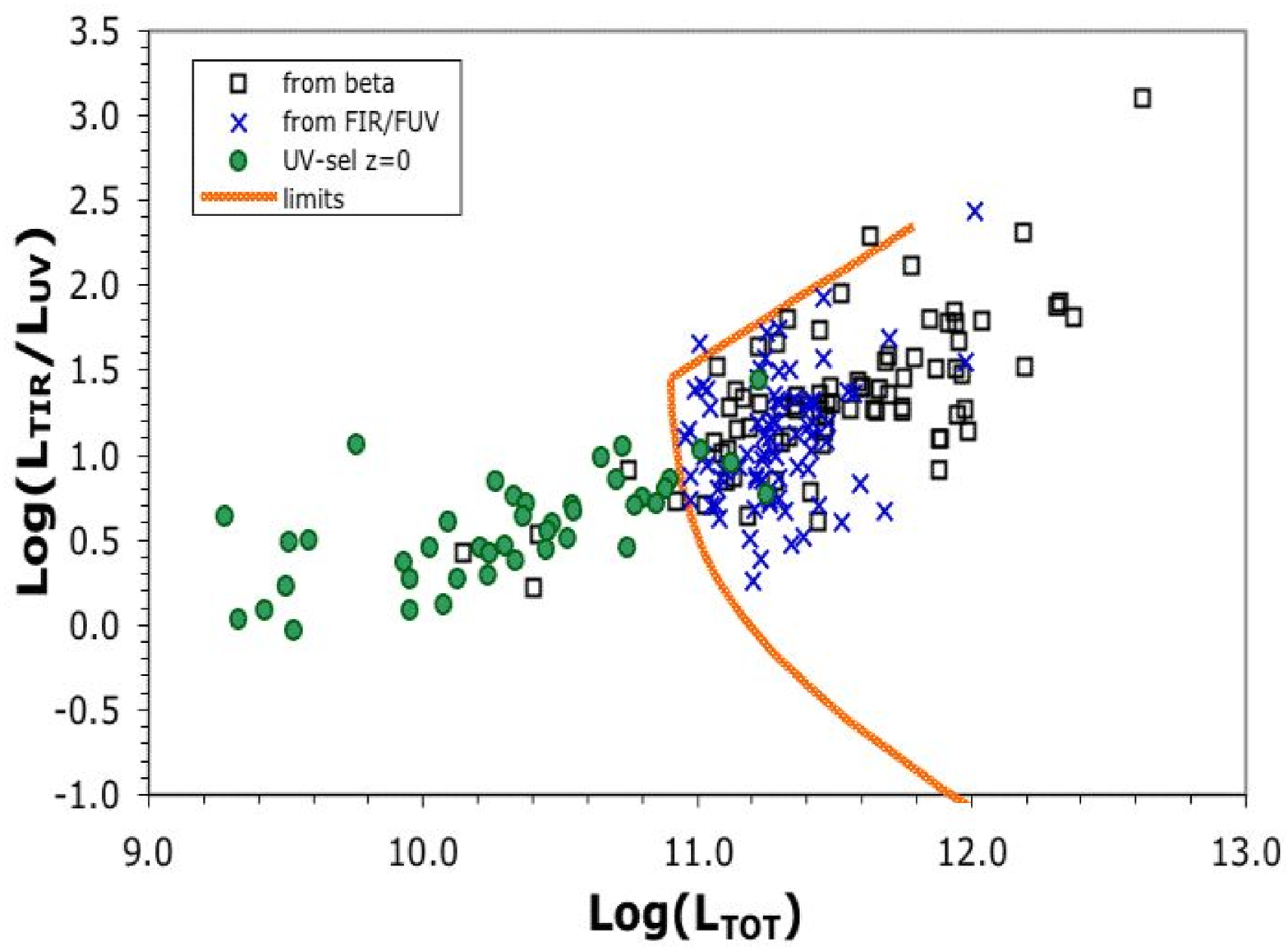,width=8cm}  
\end{center}
\vspace*{0.25cm}  
\caption{a) Blue (open boxes) and red (filled circles) symbols are the same objects but luminosities are $L_{UV}$ for the former and $L_{IR}$ for the latter. Dust attenuation strongly decreases with increasing $L_{UV}$. Part of this correlation might be due to observational limits preventing us to detect low-luminosity LBGs with low dust attenuation. The clear cut on the upper part of the box cloud cannot be due to observational biases and we do not observe UVLGs with high dust attenuations. On the other hand, the more dispersed, but well-known, increase of the dust attenuation with $L_{IR}$ is observed. b) We compare the positions of our LBG sample from our own measurements (blue crosses) based on the reliable FIR/UV ratio with the positions that would have the same objects if we apply the same method (open squares) than A\&S00. Filled green circles is the local sample of Buat et al. (2005). We observe a difference in dust attenuations : the FIR/UV ratio provides dust attenuations lower by $\sim$ 0.6 mag in average, which translate into total luminosities larger by a factor of $\sim$ 2.} 
\end{figure} 

Fig.~4a shows that the total IR luminosity correlates with the UV dust attenuation (e.g. A\&S00). On the other hand, we observe an apparent anti-correlation of $A_{FUV}$ with the UV luminosity which might be an observational bias. It is very interesting to note that we do not detect LBGs with both a high UV luminosity and a high UV dust attenuation, and this cannot be caused by observational limits. In other words, we seem to observe a population of high-$L_{UV}$ LBGs (which qualify as UVLGs) with dust attenuations similar to UV-selected galaxies in the local universe (e.g. Buat et al. 2005).

Using A\&S00 equations we estimate the IR luminosity (Fig.~4b). Globally, points estimated from the $\beta$ method are in the same region but  shifted towards the top-right part of the diagram. This method seems to overestimate $L_{IR}$ (i.e. the dust attenuation) as compared to observed FIR luminosities. The $\beta$-based average dust attenuation estimated for our sample at $z \sim 1$ is $A_{FUV} = 2.9 \pm 1.0$ while the FIR-to-UV average dust attenuation amounts to $A_{FUV} = 2.3 \pm 0.8$ (which is is fully consistent with Takeuchi, Buat \& Burgarella 2005). Both of them are marginally consistent but the net effect is that total $\beta$-based luminosities, are larger than the actual values by a factor of about 2 in average and therefore provide higher SFRs. This means that the influence of LBGs in the evolution of the star formation density might be overestimated, at least at $z \sim 1$ and the Cosmic star formation density estimated from LBGs should be decreased by a factor of 2.

This work will be described in more details in a forthcoming paper (Burgarella et al. 2005 in prep.).

\acknowledgements{We thank the Programme National Galaxies and Programme National de Cosmologie for their support. TTT has been supported by the Japan Society for the Promotion of Sciences (04/04 - 12/05).}

\vfill 
\end{document}